# Dodecahydrogen uranium: an icosahedral f-electron superatom with 26-electron shell structure

Andrii Shyichuk, Eugeniusz Zych

Faculty of Chemistry, University of Wrocław
14 F. Joliot-Curie, 50-383 Wrocław, Poland
andrii.shyichuk@uwr.edu.pl

## Abstract

A new icosahedral superatom $UH_{12}$ is reported and analyzed using post-Hartree-Fock multiconfigurational methods. A RASSCF/CASPT2/RASSI/SINGLE_ANISO workflow is employed to determine the electronic energy levels and magnetic properties. Superatomic characteristics, such as orbital topologies and shell filling order, are discussed. The high hydrogen-to-uranium ratio and potential for optically stimulated decomposition make $UH_{12}$ worth investigation in the context of hydrogen storage. The electric-dipole transition oscillator strengths of the molecule vanish, while its lowest electronic transitions fall in the THz and infrared ranges. The lowest septet manifold exhibits crystal-field splitting into a 1+3+3 pattern, as expected from icosahedral symmetry. In turn, the interlevel gaps within these crystal-field triplets are 30-150 GHz (1-5 $cm^{-1}$). The molecule is magnetically isotropic at low fields, while its levels exhibit higher-order Zeeman anisotropy that becomes significant around 2 Tesla. These properties suggest potential relevance for quantum computing, quantum communication, and quantum memory, including qubit- and qudit-based architectures.

## Introduction

Atomic orbitals – as any chemist is used to see them – are solutions of a wave equation in a spherically symmetric potential that describe the behavior of electrons. However, the potential does not have to be limited to a single nucleus. Given the right chemistry and geometry, electrons in a cluster of atoms might as well behave as not bound to a specific center, but orbiting the cluster as a whole – with certain similarities to semi-free electron gas (jellium) in conduction bands of metals (1,2). The quest for "superatoms" – clusters whose electronic shells mimic the periodic properties of isolated atoms – has largely been the playground of s and p-block elements: ease of ionization in elements like sodium or aluminum naturally leads to structures with highly delocalized molecular orbitals, even if the structures themselves are very small, namely $VNa_8$ or $Al_{13}$ (3–6). From the "magic numbers" of aluminum clusters to the noble-gas mimicry of transition-metal-doped silicon cages, the stability of these species is governed by the closure of jellium-like shells ($1S^2$, $1P^6$, $1D^{10}$,…)(1,2). However, moving into the 5f block introduces a level of complexity that challenges the standard superatom paradigm. The participation of the 5f manifold allows for high-density electronic states and relativistic effects that can either destroy or reinforce the "magic" stability(7–13).

In this work, we characterize the icosahedral $UH_{12}$ cluster, a uranium-centered "hydride" cage. At first glance, it fulfills a superatom requirement of 18 valence electrons: 6 come from U and 12 come from H. The expected shell structure is thus $1S^2 1P^6 1D^{10}$: while uranium is a 5f element, its 6d empty subshell is also chemically available. The obtained electronic properties, however, subvert this expectation: the system does not stay closed-shell, with some electrons occupying superatomic F orbitals. Using a second-order relativistic multireference approach (RASSCF/CASPT2) with a correlation-consistent relativistic core ANO-RCC-VTZP basis set(14,15), we demonstrate that $UH_{12}$ lowest manifold is a strongly muticonfigurational pseudospin septet, with a well-defined gap to the next (quintet) manifold of about 3000 $cm^{-1}$. The molecular orbitals of this system exhibit vivid atomic-like topologies (s, p, d and f).

Beyond its fundamental electronic intrigue, the exceptional stoichiometric ratio of hydrogen to uranium in $UH_{12}$ – surpassing the conventional $UH_3$ limit(10) – positions this molecule as a high-capacity candidate for solid-state hydrogen storage and a compelling model for the design of high-density hydride-moderated nuclear assemblies(16) – that is, of course, if it can be stabilized at room temperature and in practical quantities. On the other hand, the molecule exhibits effectively zero magnetic anisotropy at low fields and higher-order Zeeman anisotropy that becomes significant around 2 Tesla. The latter properties, together with the vanishing electric-dipole transition oscillator strength, suggest potential applications in quantum computing, quantum communication, and quantum memory, including qubit- and qudit-based architectures(17).

## Calculations

The calculations began with a perfect icosahedron with U-H bond length of 1.7 Angstroms and ANO-RCC-VDZP basis. Once optimized, the geometry was used as a starting point for the optimization with ANO-RCC-VTZP basis. 43 orbitals were inactive. Initially, we used the RAS-SD method where RAS1 space included the 9 highest doubly-occupied orbitals, and up to 2 holes were allowed in RAS1. In a singlet state, those orbitals exhibit superatomic topologies: 2S, 5x 1D, 3x 1F. For RAS2, we have used 6 orbitals: 3 exhibited distinct bonding superatomic 2P nature, while the other three had antibond 1F character – complementary to three bonding occupied superatomic F orbitals. With such a setup, neutral $UH_{12}$ optimization reaches a stationary point in a single-root singlet-state calculation.

In order to analyze the excited states of the system, higher spin multiplicities must be considered. For the states to be allowed to interact and mix via spin-orbit coupling, the states must have the same active spaces and the same number of allowed holes in RAS1. Given the character of the orbitals and their converged occupancies from the RAS-SD calculations, we have switched to the following RASSCF setup. 43 orbitals were inactive. RAS1 space included 6 doubly occupied orbitals (2S, 1D), and up to 2 holes were allowed in them. For RAS2, we have used another 6 orbitals: in a singlet state, 3 are occupied and have distinct superatomic 1F character, while the other 3 are empty and exhibit a clear superatomic 2P nature. In a septet state, these six orbitals maintain their character and are singly-occupied. The latter 6 orbitals were subject to full configuration interaction, but were also statically correlated with the lower semi-core orbitals. With 6 electrons in 6 orbitals, there can be between 0 and 6 unpaired electrons corresponding to spin multiplicities of 1, 3, 5 and 7. In all of those cases, these 6 orbitals are still capable of accepting up to two electrons from RAS1. Such a setup provides fewer configuration state functions than the RAS-SD one and results in much more manageable CASPT2 and RASSI calculations.

While the initial RAS-SD calculation used no symmetry and yielded an almost perfect icosahedral geometry, we have applied the highest available $D_{2h}$ symmetry in the subsequent calculations. With that, the model geometry optimizes into an icosahedron in a single-root calculation, state symmetry $A_g$ and total spin multiplicity of 1. Depending on the starting geometry, two stationary points can be reached, with bond length of 1.735 Å and 2.073 Å. Spin multiplicity 7 results in bond length 1.976726 Å, referred below as “1.977”, Table 1). Optimizing in other symmetries or spin multiplicities 3 and 5 sends the molecule on various dissociation pathways that are not explicitly discussed here. However, many correspond to formation of $H_2$ molecules with the following dissociation of the latter from the uranium center. Such behavior suggests that photoinduced or otherwise externally driven hydrogen release by this compound could be worth investigating. Considering the above, below we analyze the results for to the $D_{2h}$ septet geometry of 1.977 Å bond length. While this highly symmetric configuration serves as an excellent structural probe for the electronic shell structure, molecular orbitals and other properties, the full thermodynamic stability, potential symmetry-lowering distortions, and dissociation pathways remain an open avenue for future investigation.

All of the calculations were performed in OpenMolcas (18–20). The following modules were utilized consequently: SEWARD (integrals)(21–23), RASSCF (restricted active space self-consistent field (24–27)), CASPT2 (complete active space second-order perturbation theory, perturbative dynamic correlation correction(28–31)) and RASSI (RAS state interaction, mixing of states of different spin and symmetries (32,33)), SINGLE_ANSIO (magnetic properties (34–36)). Second-order relativistic Douglas-Kroll-Hess integrals (37–41) were utilized.

## Results and discussion

### *Oxidation states and naming*

On the one hand, H is more electronegative than U (2.2 and 1.38, respectively (42)), and hence the $UH_{12}$ molecule can be viewed as uranium(XII) dodecahydride. The +12 oxidation state of uranium, however, looks absurd. Alternatively, the molecule can be viewed as uranium(VI) dodecahydride(–0.5). Given the orbital character described below, the molecule is highly covalent and clearly not ionic. Mulliken populaions on uranium (–0.2426) and hydrogen (0.0202) support this conclusion. Uranium having a negative partial charge places this compound even further away from metal hydrides. Hence, we propose to call this molecule “dodecahydrogen uranium”.

Table 1. Cartesian x, y, z of the irreducible atoms of the 1.977 Å $D_{2h}$ geometry.

| | | | |
|---|---|---|---|
| U1 | 0 | 0 | 0 |
| H2 | 0 | 1.03922636 | 1.68150357 |
| H3 | 1.03922636 | 1.68150357 | 0 |
| H4 | 1.68150357 | 0 | 1.03922636 |

### *Molecular orbitals*

In Table 2, molecular orbitals (MOs) of the 1.977 Å $UH_{12}$ geometry are listed. The orbital numbering is specific to each irreducible representation. The displayed orbital energies of the active orbitals have been obtained (as per OpenMolcas manual) by diagonalizing the sub-blocks of the average density matrix corresponding to the different RAS orbital spaces, and thereby are named “pseudo-natural orbitals”. Both those values and the dominant atomic orbital contributions allow the superatomic shells to be assigned to these MOs. Orbitals of non-double occupation correspond to the active space.

Table 2. Selected occupied molecular orbitals of $UH_{12}$ (RASSCF, $D_{2h}$, bond length 1.977 Å), with up to three most contributing atomic orbitals and their MO coefficients.
Orbital energies from RASSCF population analysis.

| MO # | MO E | Occupancy | Irrep | Contributor 1 | | Contributor 2 | | Contributor 3 | | SA shell |
|---|---|---|---|---|---|---|---|---|---|---|
| 10 | -4.42 | 2 | $A_g$ | U1 5d0 | 0.84 | U1 5d2+ | 0.54 | | | -- |
| 11 | -4.42 | 2 | $A_g$ | U1 5d2+ | 0.84 | U1 5d0 | 0.54 | | | -- |
| 3 | -4.42 | 2 | $B_{1g}$ | U1 5d2- | 1 | | | | | -- |
| 3 | -4.42 | 2 | $B_{2g}$ | U1 5d1+ | 1 | | | | | -- |
| 3 | -4.42 | 2 | $B_{3g}$ | U1 5d1- | 1 | | | | | -- |
| 12 | -2.29 | 2 | $A_g$ | U1 6s | 0.92 | U1 7s | 0.1 | | | 1S |
| 7 | -1.24 | 2 | $B_{1u}$ | U1 6pz | 0.91 | H2 1s | 0.11 | | | 1P |
| 7 | -1.24 | 2 | $B_{2u}$ | U1 6py | 0.91 | H3 1s | 0.11 | | | 1P |
| 7 | -1.24 | 2 | $B_{3u}$ | U1 6px | 0.91 | H4 1s | 0.11 | | | 1P |
| 13 | -0.56 | 1.95 | $A_g$ | U1 6s | 0.41 | H4 1s | 0.36 | H3 1s | 0.36 | 2S |
| 14 | -0.41 | 1.92 | $A_g$ | U1 6d0 | 0.5 | H3 1s | 0.46 | H2 1s | 0.42 | 1D |
| 15 | -0.41 | 1.92 | $A_g$ | H4 1s | 0.51 | U1 6d2+ | 0.5 | H2 1s | 0.29 | 1D |
| 4 | -0.41 | 1.92 | $B_{1g}$ | H3 1s | 0.63 | U1 6d2- | 0.52 | | | 1D |
| 4 | -0.41 | 1.92 | $B_{2g}$ | H4 1s | 0.63 | U1 6d1+ | 0.52 | | | 1D |
| 4 | -0.41 | 1.92 | $B_{3g}$ | H2 1s | 0.63 | U1 6d1- | 0.52 | | | 1D |
| 8 | -0.29 | 1.1 | $B_{1u}$ | H2 1s | 0.66 | U1 6pz | 0.45 | H4 1s | 0.41 | 2P |
| 8 | -0.29 | 1.1 | $B_{2u}$ | H3 1s | 0.66 | U1 6py | 0.45 | H2 1s | 0.41 | 2P |
| 8 | -0.29 | 1.1 | $B_{3u}$ | H4 1s | 0.66 | U1 6px | 0.45 | H3 1s | 0.41 | 2P |
| 9 | -0.25 | 1.05 | $B_{1u}$ | H4 1s | 0.68 | U1 5f2+ | 0.53 | H2 1s | 0.42 | 1F |
| 9 | -0.25 | 1.05 | $B_{2u}$ | H2 1s | 0.68 | U1 5f1- | 0.6 | H3 1s | 0.42 | 1F |
| 9 | -0.25 | 1.05 | $B_{3u}$ | H3 1s | 0.68 | U1 5f3+ | 0.56 | H4 1s | 0.42 | 1F |

From Figure 1a, it is clear that the spin density in the septet state (without spin-orbit coupling) is distributed over both the uranium and hydrogen centers, without a clear preference for either. The delocalization of spin density over both elements, together with the lobes of the uranium-centered fragment pointing toward the hydrogen atoms, supports the conclusion that the bonding in the molecule has a highly covalent character. The orbitals (Fig. 2b-2i) exhibited bonding superatomic character. The overall topology is clearly similar to textbook hydrogen-like atomic orbitals, and yet those orbitals span over the whole molecule. Even the 1S orbital in Fig. 1b is in fact bonding and has contributions from hydrogen 1s atomic orbitals. However, due to the compromise isosurface value, this orbital looks spherical. Its counterpart 1P has visible lobes at hydrogens (Fig. 1d), similar could be seen in 1S if the surface was larger. Table 2 and Figure 1 show the dual nature of these MOs. On the one hand, those are clearly predominantly atomic orbitals – uranium 6s and 6p. On the other hand, these atomic orbitals overlap constructively with hydrogen 1s and have the corresponding shapes. Thus, one might view them as the lowest superatomic, and then the shell sequence is $1S^2$, $1P^6$, $2S^2$, $1D^{10}$, $2P^3$, $1F^3$ – quite close to the classical jellium model. Note that 2P and 1F orbitals are heavily mixed and exhibit similar energies. The 2S and 1D orbitals are also close in energy, and both actively participate in the configuration interaction expansion (similar to the behavior of the 4s and 3d orbitals in light transition metals, or the 7s and 6d orbitals in actinides). Consequently, this deviation from the perfect jellium model ordering should not be viewed as an argument against the overall superatomic character of the system. Comparing the 1S (Fig. 1b) and 1P (Fig. 1d) to 2S (Fig. 1c) and 2P (Fig. 1g), the lower two orbitals clearly have one radial node fewer – which is the argument to count them as superatomic 1S and 1P despite their predominant uranium atomic 6s/6p character and their semicore-like values of energies (Table 2).

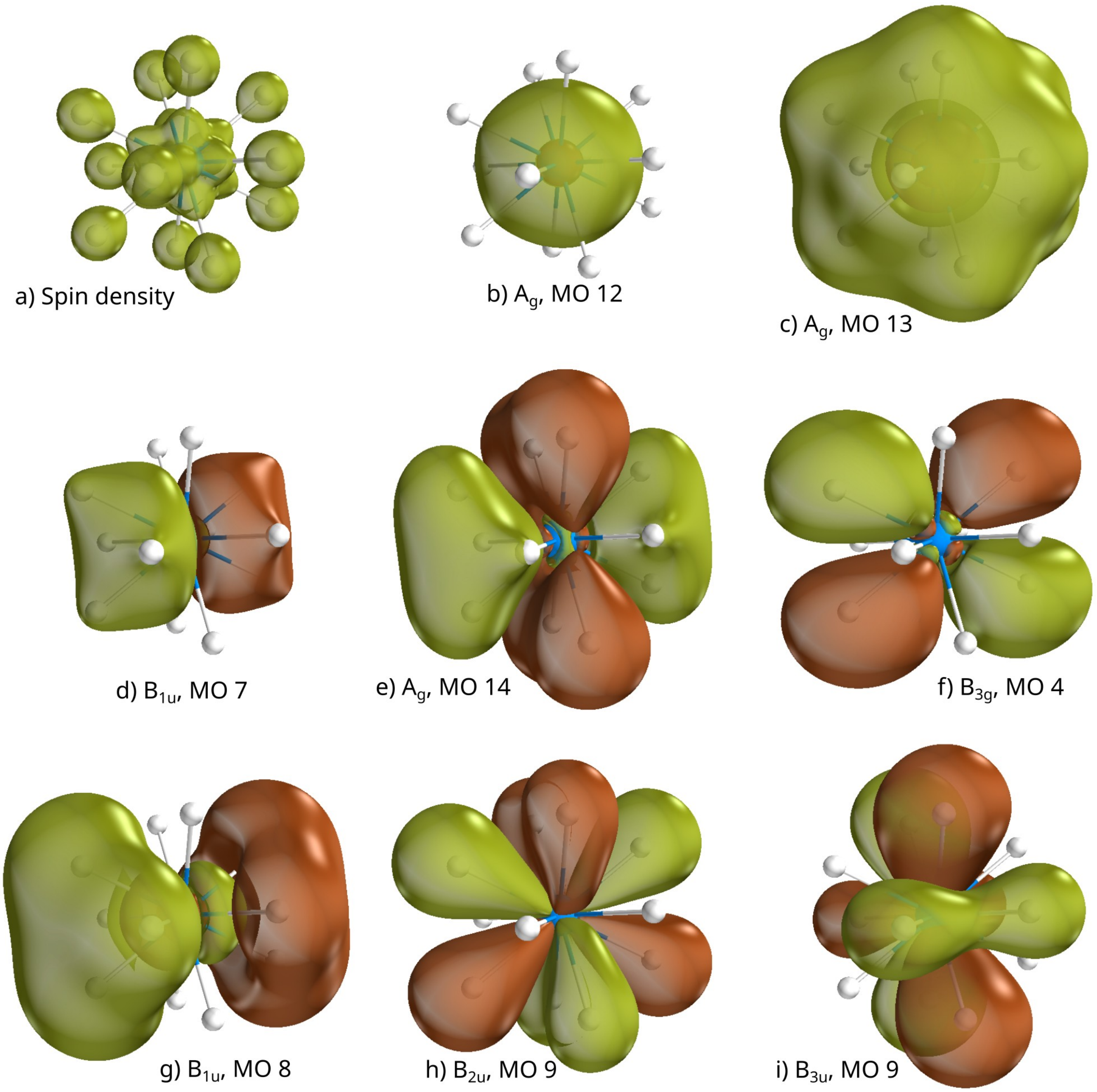

Figure 1. Total spin density (a) and selected molecular orbitals of $UH_{12}$ that illustrate basic topologies: superatomic (SA) 1S / U 6s (b), SA 1P / U 6 (d), SA 2S (c), SA 1D (e,f), SA 2P (g), SA 1F (h, i).

Hence, this dodecahydrogen uranium system exhibits a superatomic orbital structure extending into the 5f-derived manifold, with a 26-electron superatomic description that resembles conventional jellium shell ordering. From the orbital standpoint, in its high-spin form with 6 unpaired electrons, 10 orbitals are almost-doubly occupied and 6 are almost-singly occupied. In the low-spin state, all electrons are paired and occupy 13 orbitals: $1S^2$, $1P^6$, $2S^2$, $1D^{10}$, $1F^6$. Noteworthy, this sequence also deviates from the 32-electron rule, for which the jellium shell order is 1S, 1P, 1D, 1F (43,44).

The system was treated in the highest symmetry available to the code, namely $D_{2h}$. The resulting states and orbitals of even/*gerade* $B_{1g}$, $B_{2g}$, $B_{3g}$ are subduced components of the triply degenerate $T_{1g}$ or $T_{2g}$ irreducible representations of the full $I_h$ point group. The same is true for the odd/*ungerade* representations. Specifically, the orbitals shown in Fig. 1d,f-i each have two more identical counterparts in different representations. Correspondingly, CASPT2 states are triply degenerate across the $B_{1-3}$ symmetries.

Comparing $UH_{12}$ to other actinide clusters with 12 ligands such as Bi and Sb (45,46), one can notice that the latter do not exhibit such distinct atomic-like orbital shapes. In actinide clusters with 14 Au or 14 Ag atoms, the 5f orbitals stay non-valence (44) and non-superatomic, contrary to the system presented here.

*Multiconfigurational character*

In the RASSCF single-root septet calculation on the 1.977 Å geometry, the most-closed-shell configuration state function contribution weight is 77%. In such a configuration, the 2S and 1D orbitals are exactly doubly occupied. The rest of the configuration interaction expansion corresponds to various flavors of one- and two-electron excitations from the RAS1 space (the 2S and 1D) into the half-filled superatomic 2P and 1F orbitals. This result indicates that the multiconfigurational character of the molecule is rather strong. The spectrum of moderately significant excited contributions with CI coefficients larger than 0.05 is quite rich. However, such is the orbital-based picture, which lacks corrections for the dynamic correlation (CASPT2) and spin-orbit coupling. Once those are included, the corresponding change is dramatic. Triplet states turn out to be the lowest at the RMS-CASPT2 level of theory (rotated multistate CASPT2)(47). Spin-orbit coupling mixes the states of different spins and symmetries, and as a result, the spin is not a good quantum number of this system.

In Figure 2, CASPT2 energy levels are visualized. Four values of spin multiplicity and four irreducible representations are shown. Showing $B_{2g/2u}$ and $B_{3g/3u}$ levels would be redundant. To get a good representation of the lowest spin-orbit manifolds and keep the calculation size manageable, not all CASPT2 levels were passed to the RASSI state-interaction calculation. Only states under specific values of energy were selected, and the convergence of the RASSI result in respect to that value was tested. The detailed list of numbers of states is provided in Table 3 below. As can be seen in Fig. 2, state-interaction and spin-orbit coupling result in some decrease of the lowest state energy and much larger splitting of the treated subspace. As more levels are included, a distinct gap is formed, and two lowest manifolds crystallize: 7 levels with pseudospin value 2.9 (~3, septet) and 5 levels of pseudospin 1.8 (~2, quintet) (in calculations with the selection thresholds of .668 and .6627 Hartree). For example, in the calculation with .674 Hartree threshold – the one considered not yet complete – the first manifold was already formed, while the second one comprised 5 states of pseudospin value 3.2, too few states for a septet. Once the two manifolds were established, they were subjected to SINGLE_ANISO calculation of the magnetic properties, with the corresponding pseudospin values of 3 and 2.

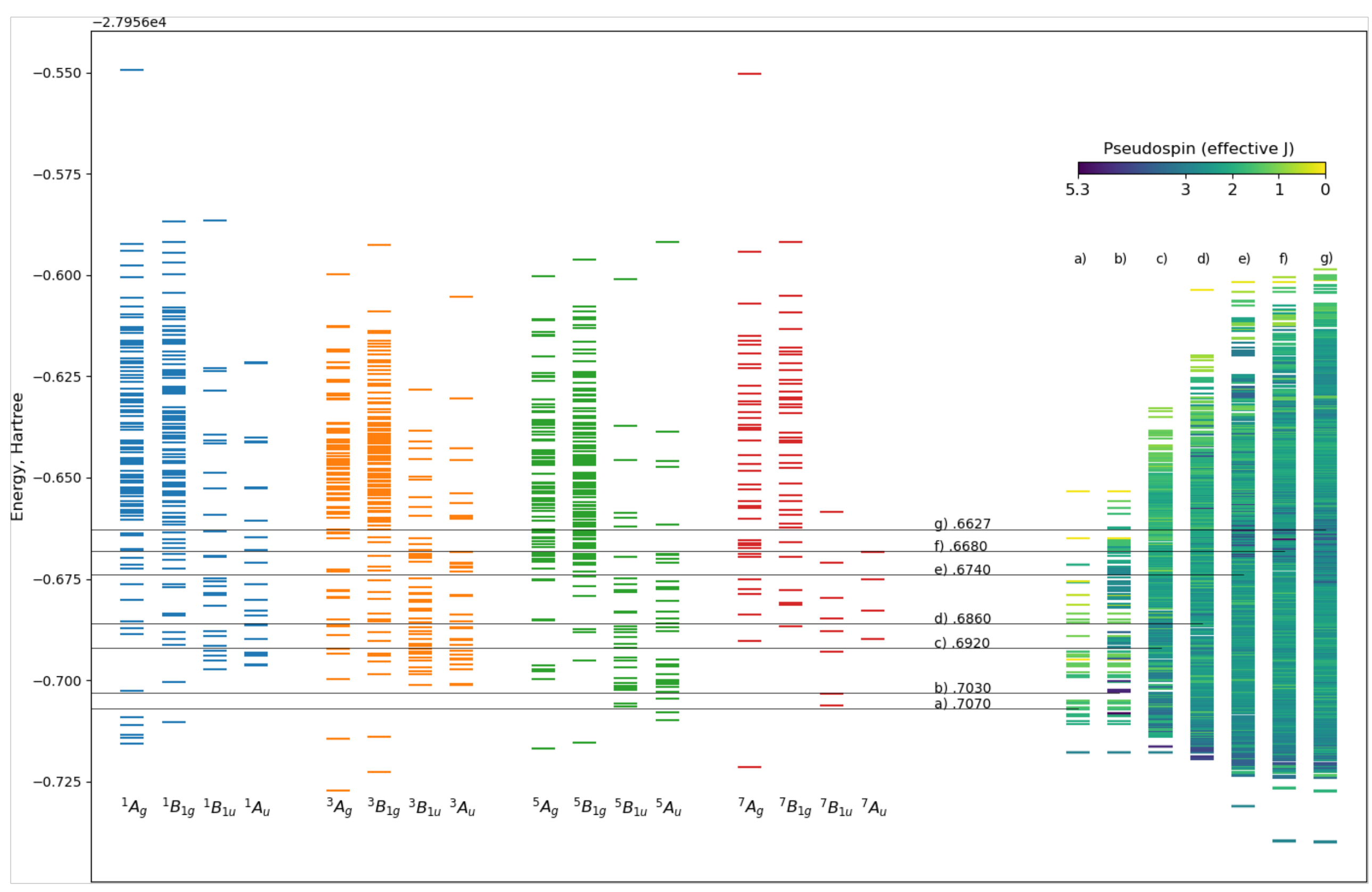

Figure 2. RMS-CASPT2 states (left) and the corresponding mixed spin-orbit states (right), as a function of selection threshold indicated by horizontal lines and values a-g (see text). The effective J value (the pseudospin) is encoded by color in the right hand plot.

Table 3. Number of states in CASPT2 calculation (full) and the subsequent RASSI calculations. The energy threshold is –27956–*x* Hartree.

| $x$ | 2S+1 | $A_g$ | $B_{1g}$ | $B_{2g}$ | $B_{3g}$ | $B_{1u}$ | $B_{2u}$ | $B_{3u}$ | $A_u$ | $x$ | 2S+1 | $A_g$ | $B_{1g}$ | $B_{2g}$ | $B_{3g}$ | $B_{1u}$ | $B_{2u}$ | $B_{3u}$ |
|---|---|---|---|---|---|---|---|---|---|---|---|---|---|---|---|---|---|---|
| full | 1 | 93 | 94 | 94 | 94 | 26 | 26 | 26 | 26 | .686 | 1 | 8 | 5 | 5 | 5 | 7 | 7 | 7 |
| | 3 | 100 | 100 | 100 | 100 | 45 | 45 | 45 | 45 | | 3 | 10 | 8 | 8 | 8 | 17 | 17 | 17 |
| | 5 | 100 | 100 | 100 | 100 | 31 | 31 | 31 | 31 | | 5 | 6 | 4 | 4 | 4 | 17 | 17 | 17 |
| | 7 | 41 | 37 | 37 | 37 | 8 | 8 | 8 | 6 | | 7 | 2 | 1 | 1 | 1 | 4 | 4 | 4 |
| .6627 | 1 | 18 | 17 | 17 | 17 | 16 | 16 | 16 | 18 | .692 | 1 | 6 | 2 | 2 | 2 | 4 | 4 | 4 |
| | 3 | 22 | 20 | 20 | 20 | 35 | 35 | 35 | 33 | | 3 | 7 | 6 | 6 | 6 | 10 | 10 | 10 |
| | 5 | 30 | 25 | 25 | 25 | 25 | 25 | 25 | 25 | | 5 | 6 | 2 | 2 | 2 | 12 | 12 | 12 |
| | 7 | 12 | 6 | 6 | 6 | 7 | 7 | 7 | 6 | | 7 | 1 | 0 | 0 | 0 | 3 | 3 | 3 |
| .668 | 1 | 14 | 12 | 12 | 12 | 15 | 15 | 15 | 15 | .703 | 1 | 5 | 1 | 1 | 1 | 0 | 0 | 0 |
| | 3 | 18 | 17 | 17 | 17 | 32 | 32 | 32 | 33 | | 3 | 3 | 2 | 2 | 2 | 0 | 0 | 0 |
| | 5 | 18 | 15 | 15 | 15 | 25 | 25 | 25 | 25 | | 5 | 2 | 1 | 1 | 1 | 3 | 3 | 3 |
| | 7 | 8 | 5 | 5 | 5 | 7 | 7 | 7 | 6 | | 7 | 1 | 0 | 0 | 0 | 2 | 2 | 2 |
| .674 | 1 | 11 | 9 | 9 | 9 | 13 | 13 | 13 | 14 | .707 | 1 | 5 | 1 | 1 | 1 | 0 | 0 | 0 |
| | 3 | 15 | 14 | 14 | 14 | 27 | 27 | 27 | 24 | | 3 | 3 | 2 | 2 | 2 | 0 | 0 | 0 |
| | 5 | 10 | 7 | 7 | 7 | 23 | 23 | 23 | 21 | | 5 | 2 | 1 | 1 | 1 | 0 | 0 | 0 |
| | 7 | 6 | 4 | 4 | 4 | 6 | 6 | 6 | 0 | | 7 | 1 | 0 | 0 | 0 | 0 | 0 | 0 |

*Energy levels and magnetic properties*

The two sets of levels corresponding to the selection thresholds of .6627 and .668 Ha exhibited very similar electronic properties. Both are kept to highlight the dependence of the properties on the subjective parameter such as state selection. The lowest (septet) manifold in both systems is characterized by the same pattern of states that lie at slightly different energies with respect to each other, while the gaps between them stay consistent.

As shown in Figure 3, the lowest manifold spans from 0 to about 70 $cm^{-1}$, the values are listed in Table 4. The seven levels are grouped into a subsets of a singlet, and two triplets, in line with a septet splitting under icosahedral crystal field into a $A_g$+$T_{1g}$+$T_{2g}$ representations. Below, these are referred to as "crystal field triplets" to avoid confusion with spin multiplicity triplets mentioned above. All seven levels exhibit a similar kind of spin and representation mixing, and are comprised of about 60% $^7A_g$ states and 40% of $^5B_{1g/2g/3g}$ states – note the different spin multiplicities. The next manifold – the quintet – is composed of roughly 30% $^5A_g$ states and 70% of $^5B_{1g/2g/3g}$.

The septet manifold provides several low-energy transitions that happen to lie in GHz range, namely ~30, 40 and 70 $cm^{-1}$, or 900, 1200 and 2100 GHz. On the one hand, such energy gaps are too large for the 1-100 GHz required for qubit transitions (17). On the other hand, the gaps inside the crystal field triplet submanifolds are in the range of 1-5 $cm^{-1}$ – 30 to 150 GHz. As will be shown below, these gaps can be fine-tuned by application of an external magnetic field. The large 3000 $cm^{-1}$ gap separating this manifold from the next one prevents any spurious state interaction in the manipulation and readout range. These properties make $UH_{12}$ a promising theoretical platform for further investigation in the context of quantum computing and quantum information technologies.

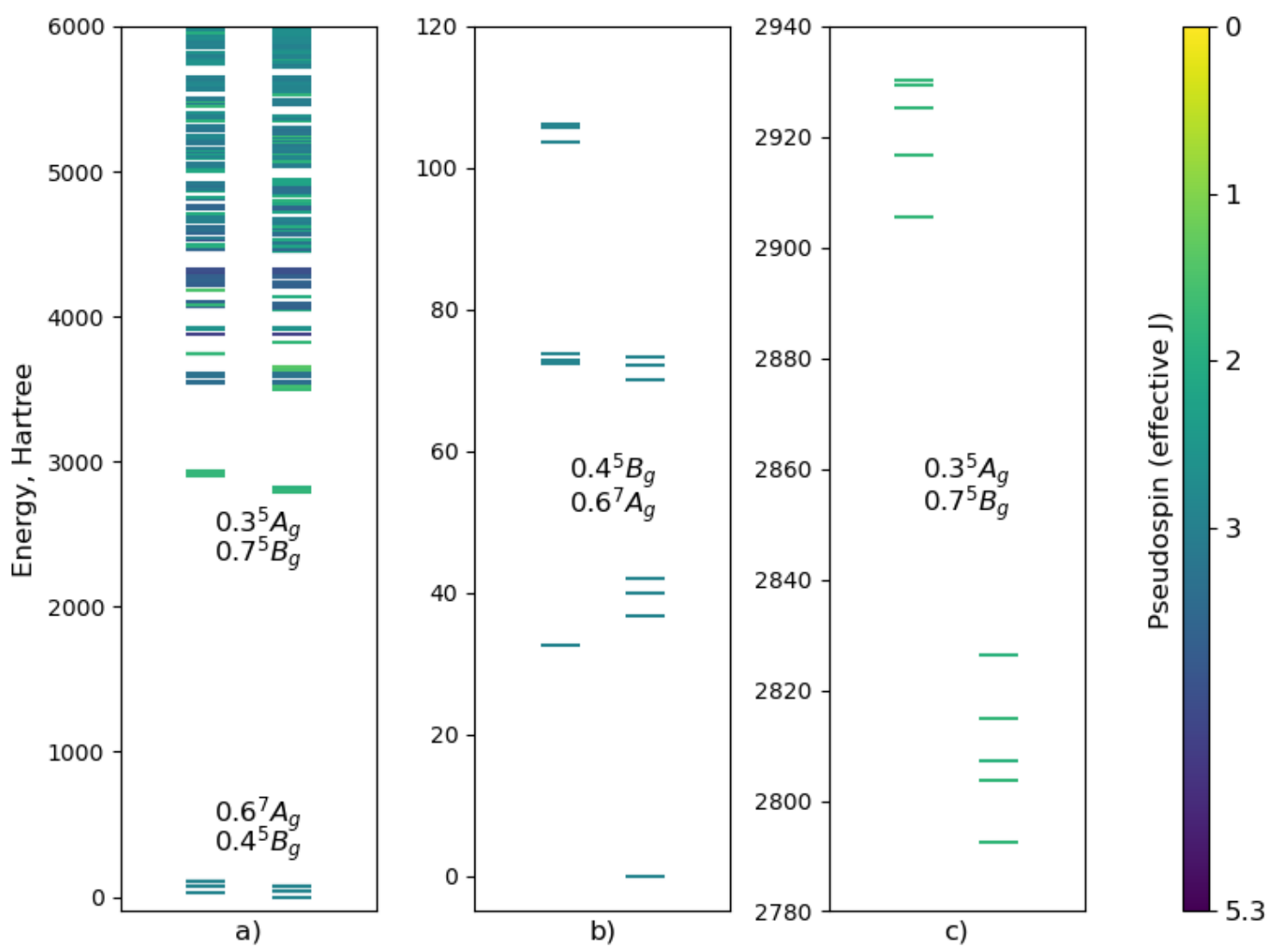

Figure 3. The energy levels from the two calculations considered complete, under 6000 $cm^{-1}$ (a), under 80 $cm^{-1}$ (b, the lowest manifold, septet), and 2780-2940 $cm^{-1}$ (c, the second manifold, quintet)

In Table 4, the most important magnetic characteristics of the two manifolds are summarized, two sets for the two threshold values. The sets are not identical, but the differences are not large either, indicating a reasonable convergence of the results with respect to the number of selected RMS-CASPT2 states.

Table 4. Energy levels and magnetic properties of the two lowest manifolds of $UH_{12}$ superatom, in the units of $cm^{-1}$.

| Energy levels | | | | Property | Property component | | | | | |
|---|---|---|---|---|---|---|---|---|---|---|
| .6680 | | .6627 | | | .6680 | | | .6627 | | |
| As is | Shifted | As is | Shifted | | x | y | z | x | y | z |
| Septet, $J_{eff.}$ = 3 | | | | | $D_A$ = 0.297, $E_A$ = -0.008 | | | $D_A$ = 0.169, $E_A$ = -0.018 | | |
| 0.0 | | 0.0 | | | | | | | | |
| 39.5 | | 36.7 | | g | 1.785 | 1.785 | 1.785 | 1.781 | 1.781 | 1.782 |
| 40.2 | | 40.1 | | μ | -5.356 | -5.356 | -5.357 | -5.344 | -5.345 | -5.349 |
| 40.3 | | 42.1 | | L | 0.338 | 0.338 | 0.338 | 0.337 | 0.337 | 0.337 |
| 70.4 | | 70.1 | | S | 2.506 | 2.506 | 2.506 | 2.501 | 2.501 | 2.503 |
| 72.3 | | 72.1 | | D | 0.198 | -0.107 | -0.091 | 0.113 | -0.074 | -0.039 |
| 72.6 | | 73.4 | | | | | | | | |
| Quintet, $J_{eff.}$ = 2 | | | | | $D_A$ = 2.248, $E_A$ = 2.233 | | | $D_A$ = 3.660, $E_A$ = 2.179 | | |
| 2891.7 | 0.0 | 2792.4 | 0.0 | g | 1.245 | 1.245 | 1.241 | 1.230 | 1.222 | 1.206 |
| 2910.4 | 18.7 | 2803.7 | 11.3 | μ | -2.516 | -2.525 | -2.504 | -2.489 | -2.486 | -2.449 |
| 2929.3 | 37.6 | 2807.3 | 14.9 | L | 0.227 | 0.227 | 0.228 | 0.212 | 0.212 | 0.215 |
| 2941.8 | 50.1 | 2814.8 | 22.4 | S | 1.143 | 1.148 | 1.137 | 1.137 | 1.136 | 1.116 |
| 2942.1 | 50.4 | 2826.2 | 33.8 | D | 1.484 | -2.982 | 1.499 | 0.959 | -3.399 | 2.440 |

The septet manifold is characterized by a high degree of magnetic isotropy. The *x*, *y* and *z* components of core properties such as the g-tensor, magnetic moment (μ), orbital angular momentum, and spin are effectively identical. For the latter three, the off-diagonal elements are zero. The D tensor stands out slightly and has different component values. The anisotropy parameters $D_A$ and $E_A$ are very small, indicating the effective isotropic property. As per group theory, in icosahedral symmetry both axial and rhombic anisotropy parameters ($D_A$ and $E_A$ respectively) should be exactly zero (48). Here, despite constraints, small non-zero values are observed. We view them as an effective error gauge of the approach, and attribute them to numerical noise, incompleteness with respect to full configuration-interaction expansion, and other limitations of the CASPT2/RASSI methodology. The corresponding magnetic moment μ is effectively the same in all directions, indicating the isotropy. The principal magnetization axes correspond to the icosahedron geometry principal axes. The quintet manifold does exhibit a certain degree of anisotropy and slightly higher values of $D_A$ and $E_A$ – however, those are still low and can be considered as effectively zero (Table 4). The system is characterized by low orbital angular momentum. Angular momentum quenching is considered characteristic for superatomic systems (49).

The question of crystal field triplet nondegeneracy and the observed tiny anisotropies can be interpreted twofold. On the one hand, the system is icosahedral, and must exhibit a degeneracy in the crystal field triplets, as well as exactly zero anisotropy. Hence, the deviations should be attributed to broadly understood artifacts aggregating pure numerical errors, basis set, and a stack of active space incompleteness at RASSCF, CASPT2, and RASSI. If so, the observed values of several $cm^{-1}$ should be considered good result. On the other hand, the system was calculated in $D_{2h}$ Abelian symmetry, in which each of the crystal field triplet levels falls in the different $B_{1,2,3}$ irreducible representations and the degeneracy is naturally lifted. That would make the observed differences physical – or at least more physical – than pure artifact. That assumption is easy to test, and to do so we ran the same calculation on a purely icosahedral geometry of the same bond length as the optimized one, 1.976726 Å. At RMS-CASPT2, the energy differences between the two systems were not systematic: for some values of spin and state symmetry, the perfect geometry corresponded to a less energetic lowest state, and for others – the optimized one. At RASSI level, the optimized geometry lowest state was 0.03 mHa or 7.4 $cm^{-1}$ more energetic. Hence, technically, ideal icosahedral geometry is preferred, but with an energy difference so minuscule that it is rather inconclusive. The energy differences between the septet levels were slightly larger for the optimized geometry. In the icosahedral one, the degeneracy is also lifted – be that as it may, it is still a $D_{2h}$ calculation from a wavefunction point of view. However, in the optimized geometry, the energy differences between crystal field triplet levels are on average about 1 $cm^{-1}$ larger than in the perfectly icosahedral one, indicating that the differences originate – at least in part – from the ever-so-slight distortion of the icosahedral geometry into geometrical $D_{2h}$.

Finally, from a statistical point of view, there are reasons to believe that nature has tendency to lower-symmetry polyhedra (50). At temperatures higher than absolute zero this molecule will – as any other molecule – undergo fluctuations of shape leading to distortions of the perfect icosahedral shape and, yet again, to lifting of the degeneracy. If an external magnetic field is applied, the system symmetry will be reduced by the field. Summarizing, we tend not to discard the observed deviations as pure artifacts.

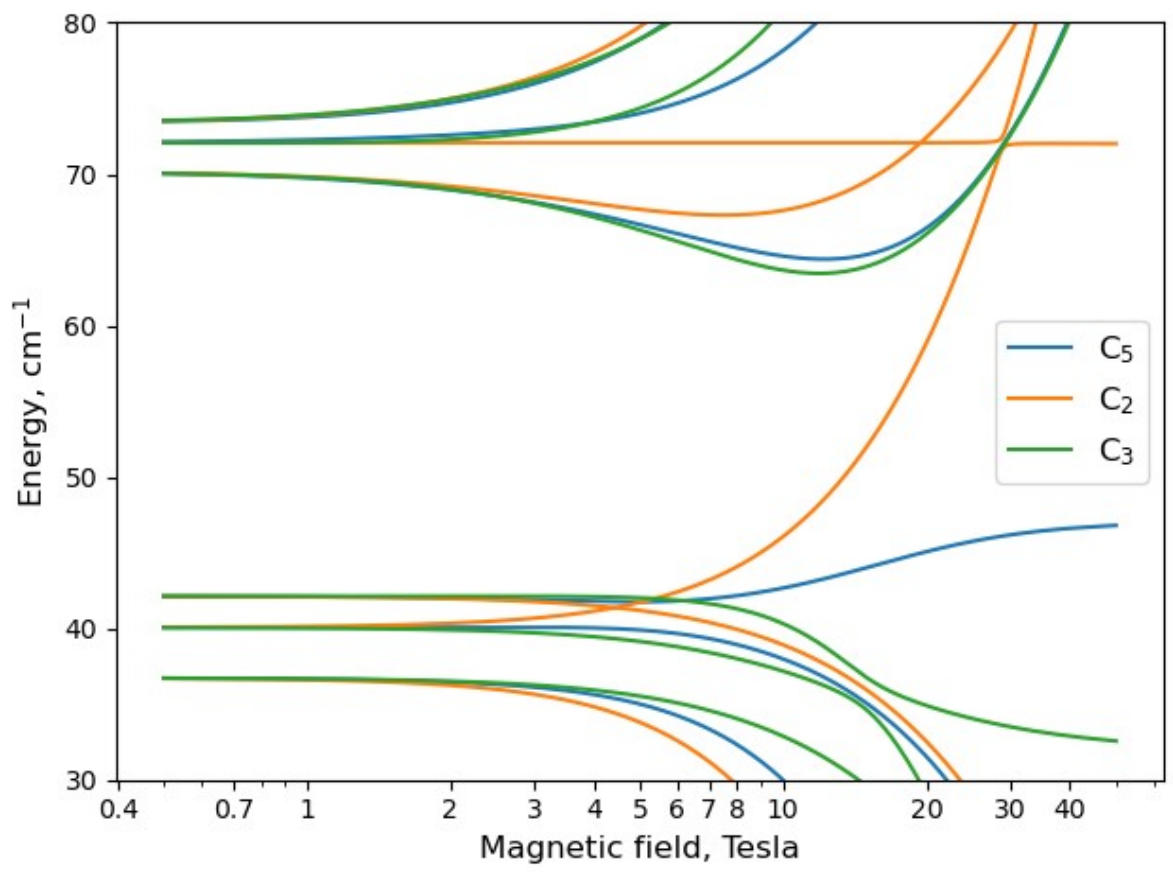

Figure 4. Axis-dependent Zeeman effect in the .6627 system.

The dependence of the energy levels on magnetic field has been calculated as well, using SINGLE_ANISO, up to 50 Tesla. Three different axes of the molecule were studied: $C_5$ (along the U-H bond), $C_2$ (along principal axis *z*) and $C_3$ (in between thee U-H bonds, or the 111 direction). As seen in Fig. 4, at lower field values (below about 2 Tesla) the levels exhibit relatively weak Zeeman effect, as expected given the small values of the anisotropy parameters and high symmetry: lower rank Zeeman terms are suppressed. However, at higher values of the field, the Zeeman effect becomes stronger and distinctly dependent on the field direction. From a practical point of view, the observed Zeeman effect might be used to fine-tune transitions between the levels to the specific GHz frequency range of the experimental setup – already at moderate fields of 2-4 Tesla.

While the primary interest of this system as a qubit lies in its crystal field triplet states, application of strong magnetic fields can change transition frequencies significantly, thus achieving a system with more than 3 levels of similar transition energies. Having individual (non-equal) transitions between consecutive levels allows for selective manipulation of states and readout. In other words, this molecule may provide flexible, adaptive, radiofrequency-addressable and controlled molecular spin states of potential relevance to molecular quantum information.

Despite the high pseudospin and intrinsic magnetic moment, the molecule does not present itself as a potential single-molecule magnet – due to the observed magnetic isotropy and, consequently, the lack of a strong axis. The symmetry of the system – either $I_h$ or $D_{2h}$ – is characterized by inversion, leading to the calculated transition dipole moment values of exactly zero. Such a property can potentially suppress electric-dipole-mediated radiative relaxation and decoherence channels (sensitivity to electric-dipole noise is a major limitation of quantum bits (17)). Application of an external magnetic field should, as shown above, disregard the underlying degeneracies (if any) in the levels of the septet manifold and split it into 7 independent states. The energy gaps between them can be fine-tuned by the applied magnetic field to match the specific preparation and readout setup. Finally, the composition of uranium and hydrogen allows isotope selectivity to reduce decoherence via interaction with nuclear spins. Replacement $^{1}H$ with $^{2}H$ (deuterium) can alter and potentially reduce hyperfine coupling and nuclear-spin noise (51). Concurrently, utilizing $^{238}U$ provides a strictly zero nuclear spin at the heavy-atom core, eliminating central hyperfine splittings entirely. Such isotopic substitutions could provide an additional experimental parameter for controlling nuclear-spin-related decoherence, although the relevant hyperfine and spin-phonon interactions remain to be quantified.

## Conclusions

In this work, we describe – for the first time, to the best of our knowledge – an icosahedral superatom molecule called dodecahydrogen uranium. It comprises 26 electrons to be considered as superatomic, and the shell filling pattern of: $1S^2$, $1P^6$, $2S^2$, $1D^{10}$, $2P^3$, $1F^3$. This $D_{2h}$ 1.977 Å system is characterized by a high-spin ground state, strong multiconfigurational character and distinct spin-orbit coupling effects. The lowest manifold is a septet with an effective J value (pseudospin) of 3, followed by a 3000 $cm^{-1}$ gap separating it from the next manifold that is a quintet ($J_{eff.} = 2$). According to icosahedral crystal field, the septet is split into crystal field submanifolds: a singlet, and two triplets located ~40 and 70 $cm^{-1}$ (1.2 and 2.1 THz) higher. The transitions between the crystal field triplet levels are in the range of 30-150 GHz (1-5 $cm^{-1}$). Both lowest manifolds exhibit negligible magnetic anisotropy. The aforementioned properties make this molecule an interesting theoretical platform for investigating radiofrequency-addressable molecular spin states, with potential relevance for quantum computing and communication.

## References

1. Knight WD, Clemenger K, De Heer WA, Saunders WA, Chou MY, Cohen ML. Electronic Shell Structure and Abundances of Sodium Clusters. Phys Rev Lett. 1984 Jun 11;52(24):2141–3. doi:10.1103/PhysRevLett.52.2141
2. Rao BK, Khanna SN, Jena P. Designing New Materials Using Atomic Clusters.
3. Shibuta M, Inoue T, Kamoshida T, Eguchi T, Nakajima A. Al13− and B@Al12− superatoms on a molecularly decorated substrate. Nat Commun. 2022 Mar 14;13(1):1336. doi:10.1038/s41467-022-29034-9
4. Jia Y, Luo Z. Thirteen-atom metal clusters for genetic materials. Coordination Chemistry Reviews. 2019 Dec;400:213053. doi:10.1016/j.ccr.2019.213053
5. Jena P, Sun Q. Super Atomic Clusters: Design Rules and Potential for Building Blocks of Materials. Chem Rev. 2018 Jun 13;118(11):5755–870. doi:10.1021/acs.chemrev.7b00524
6. Reber AC, Khanna SN. Superatoms: Electronic and Geometric Effects on Reactivity. Acc Chem Res. 2017 Feb 21;50(2):255–63. doi:10.1021/acs.accounts.6b00464
7. Pyykkö P. Relativistic Theory of Atoms and Molecules III: A Bibliography 1993–1999 [Internet]. Vol. 76. Berlin, Heidelberg: Springer Berlin Heidelberg; 2000 [cited 2026 Jun 5]. (Berthier G, Fischer H, Fukui K, Hall GG, Hinze J, Jortner J, et al., editors. Lecture Notes in Chemistry). Available from: http://link.springer.com/10.1007/978-3-642-51885-0 doi:10.1007/978-3-642-51885-0
8. Pyykko P. Relativistic effects in structural chemistry. Chem Rev. 1988 May 1;88(3):563–94. doi:10.1021/cr00085a006
9. Drummond Turnbull R, Bell NL. f-Block hydride complexes – synthesis, structure and reactivity. Dalton Trans. 2024;53(31):12814–36. doi:10.1039/D4DT00776J
10. Kruglov IA, Kvashnin AG, Goncharov AF, Oganov AR, Lobanov SS, Holtgrewe N, et al. Uranium polyhydrides at moderate pressures: Prediction, synthesis, and expected superconductivity. Sci Adv. 2018 Oct 5;4(10):eaat9776. doi:10.1126/sciadv.aat9776
11. Gagliardi L, Roos BO. Quantum chemical calculations show that the uranium molecule U2 has a quintuple bond. Nature. 2005 Feb;433(7028):848–51. doi:10.1038/nature03249
12. Boronski JT, Seed JA, Hunger D, Woodward AW, Van Slageren J, Wooles AJ, et al. A crystalline tri-thorium cluster with σ-aromatic metal–metal bonding. Nature. 2021 Oct 7;598(7879):72–5. doi:10.1038/s41586-021-03888-3
13. Beltran-Leiva MJ, Batista ER, Yang P. Unlocking Novel δ and φ Bonding Modes in Actinides via Oxidation State Control. JACS Au. 2025 Apr 28;5(4):1746–59. doi:10.1021/jacsau.4c01277
14. Widmark PO. Density matrix averaged atomic natural orbital (ANO) basis sets for correlated molecular wave functions.
15. Roos BO, Lindh R, Malmqvist PÅ, Veryazov V, Widmark PO. New relativistic ANO basis sets for actinide atoms. Chemical Physics Letters. 2005 Jun;409(4–6):295–9. doi:10.1016/j.cplett.2005.05.011
16. Hypes-Mayfield V, Kubic W, Dogruel D, Hollis K, Willms S, Dumont JH. Uranium Bed Design Parameters for Tritium Plants Supporting Fusion Reactors. Fusion Science and Technology. 2021 Nov 17;77(7–8):836–41. doi:10.1080/15361055.2021.1883978
17. Gaita-Ariño A, Luis F, Hill S, Coronado E. Molecular spins for quantum computation. Nat Chem. 2019 Apr;11(4):301–9. doi:10.1038/s41557-019-0232-y

18. Aquilante F, Autschbach J, Baiardi A, Battaglia S, Borin VA, Chibotaru LF, et al. Modern quantum chemistry with [Open]Molcas. The Journal of Chemical Physics. 2020 Jun 7;152(21):214117. doi:10.1063/5.0004835
19. Fdez. Galván I, Vacher M, Alavi A, Angeli C, Aquilante F, Autschbach J, et al. OpenMolcas: From Source Code to Insight. J Chem Theory Comput. 2019 Nov 12;15(11):5925–64. doi:10.1021/acs.jctc.9b00532
20. Olsen J, Roos BO, Jo/rgensen P, Jensen HJAa. Determinant based configuration interaction algorithms for complete and restricted configuration interaction spaces. The Journal of Chemical Physics. 1988 Aug 15;89(4):2185–92. doi:10.1063/1.455063
21. Lindh R, Ryu U, Liu B. The reduced multiplication scheme of the Rys quadrature and new recurrence relations for auxiliary function based two-electron integral evaluation. The Journal of Chemical Physics. 1991 Oct 15;95(8):5889–97. doi:10.1063/1.461610
22. Heß BA, Marian CM, Wahlgren U, Gropen O. A mean-field spin-orbit method applicable to correlated wavefunctions. Chemical Physics Letters. 1996 Mar;251(5–6):365–71. doi:10.1016/0009-2614(96)00119-4
23. Schimmelpfennig B. AMFI, an atomic mean-field spin–orbit integral program. University of Stockholm; 1996.
24. Roos BO, Taylor PR, Sigbahn PEM. A complete active space SCF method (CASSCF) using a density matrix formulated super-CI approach. Chemical Physics. 1980 May;48(2):157–73. doi:10.1016/0301-0104(80)80045-0
25. Roos BO. The complete active space SCF method in a fock-matrix-based super-CI formulation. Int J Quantum Chem. 2009 Jun 19;18(S14):175–89. doi:10.1002/qua.560180822
26. Roos BO. The Complete Active Space Self-Consistent Field Method and its Applications in Electronic Structure Calculations. In: Lawley KP, editor. Advances in Chemical Physics [Internet]. 1st ed. Wiley; 1987 [cited 2025 Dec 29]. p. 399–445. Available from: https://onlinelibrary.wiley.com/doi/10.1002/9780470142943.ch7 doi:10.1002/9780470142943.ch7
27. Malmqvist PAake, Rendell Alistair, Roos BO. The restricted active space self-consistent-field method, implemented with a split graph unitary group approach. J Phys Chem. 1990 Jul;94(14):5477–82. doi:10.1021/j100377a011
28. Forsberg N, Malmqvist PÅ. Multiconfiguration perturbation theory with imaginary level shift. Chemical Physics Letters. 1997 Aug;274(1–3):196–204. doi:10.1016/S0009-2614(97)00669-6
29. Andersson Kerstin, Malmqvist PAake, Roos BO, Sadlej AJ, Wolinski Krzysztof. Second-order perturbation theory with a CASSCF reference function. J Phys Chem. 1990 Jul;94(14):5483–8. doi:10.1021/j100377a012
30. Andersson K, Malmqvist PÅ, Roos BO. Second-order perturbation theory with a complete active space self-consistent field reference function. The Journal of Chemical Physics. 1992 Jan 15;96(2):1218–26. doi:10.1063/1.462209
31. Finley J, Malmqvist PÅ, Roos BO, Serrano-Andrés L. The multi-state CASPT2 method. Chemical Physics Letters. 1998 May;288(2–4):299–306. doi:10.1016/S0009-2614(98)00252-8
32. Malmqvist PÅ, Roos BO. The CASSCF state interaction method. Chemical Physics Letters. 1989 Feb;155(2):189–94. doi:10.1016/0009-2614(89)85347-3
33. Malmqvist PÅ. Calculation of transition density matrices by nonunitary orbital transformations. Int J of Quantum Chemistry. 1986 Oct;30(4):479–94. doi:10.1002/qua.560300404
34. Chibotaru LF, Ungur L, Aronica C, Elmoll H, Pilet G, Luneau D. Structure, Magnetism, and Theoretical Study of a Mixed-Valence $Co^{II}_3$ $Co^{III}_4$ Heptanuclear Wheel: Lack of SMM Behavior despite Negative Magnetic Anisotropy. J Am Chem Soc. 2008 Sep 17;130(37):12445–55. doi:10.1021/ja8029416

35. Chibotaru LF, Ungur L. *Ab initio* calculation of anisotropic magnetic properties of complexes. I. Unique definition of pseudospin Hamiltonians and their derivation. The Journal of Chemical Physics. 2012 Aug 14;137(6):064112. doi:10.1063/1.4739763
36. Chibotaru LF, Ungur L, Soncini A. The Origin of Nonmagnetic Kramers Doublets in the Ground State of Dysprosium Triangles: Evidence for a Toroidal Magnetic Moment. Angew Chem Int Ed. 2008 May 19;47(22):4126–9. doi:10.1002/anie.200800283
37. Peng D, Hirao K. An arbitrary order Douglas–Kroll method with polynomial cost. The Journal of Chemical Physics. 2009 Jan 28;130(4):044102. doi:10.1063/1.3068310
38. Reiher M. Douglas–Kroll–Hess Theory: a relativistic electrons-only theory for chemistry. Theor Chem Acc. 2006 Jul 31;116(1–3):241–52. doi:10.1007/s00214-005-0003-2
39. Reiher M, Wolf A. Exact decoupling of the Dirac Hamiltonian. I. General theory. The Journal of Chemical Physics. 2004 Aug 1;121(5):2037–47. doi:10.1063/1.1768160
40. Reiher M, Wolf A. Exact decoupling of the Dirac Hamiltonian. II. The generalized Douglas–Kroll–Hess transformation up to arbitrary order. The Journal of Chemical Physics. 2004 Dec 8;121(22):10945–56. doi:10.1063/1.1818681
41. Peng D, Reiher M. Exact decoupling of the relativistic Fock operator. Theor Chem Acc. 2012 Jan;131(1):1081. doi:10.1007/s00214-011-1081-y
42. Pauling L. THE NATURE OF THE CHEMICAL BOND. IV. THE ENERGY OF SINGLE BONDS AND THE RELATIVE ELECTRONEGATIVITY OF ATOMS. J Am Chem Soc. 1932 Sep;54(9):3570–82. doi:10.1021/ja01348a011
43. Dognon JP. Theoretical insights into the chemical bonding in actinide complexes. Coordination Chemistry Reviews. 2014 May;266–267:110–22. doi:10.1016/j.ccr.2013.11.018
44. Zhao J, Du Q, Zhou S, Kumar V. Endohedrally Doped Cage Clusters. Chem Rev. 2020 Sep 9;120(17):9021–163. doi:10.1021/acs.chemrev.9b00651
45. Zhang NX, Wang CZ, Lan JH, Wu QY, Shi WQ. Actinide endohedral inter-metalloid clusters of the group 15 elements. Phys Chem Chem Phys. 2024;26(38):25069–75. doi:10.1039/D4CP02546F
46. McGrady JE, Weigend F, Dehnen S. Electronic structure and bonding in endohedral Zintl clusters. Chem Soc Rev. 2022;51(2):628–49. doi:10.1039/D1CS00775K
47. Battaglia S, Lindh R. On the role of symmetry in XDW-CASPT2. The Journal of Chemical Physics. 2021 Jan 21;154(3):034102. doi:10.1063/5.0030944
48. Cahier B, Maurice R, Bolvin H, Mallah T, Guihéry N. Tools for Predicting the Nature and Magnitude of Magnetic Anisotropy in Transition Metal Complexes: Application to Co(II) Complexes. Magnetochemistry. 2016 Aug 3;2(3):31. doi:10.3390/magnetochemistry2030031
49. Roduner E. Superatom chemistry: promising properties of near-spherical noble metal clusters. Phys Chem Chem Phys. 2018;20(37):23812–26. doi:10.1039/C8CP04651D
50. Munguba GHL, Da Silva MF, Silva FT, Urquiza-Carvalho GA, Simas AM. Thermally Distinguishable Polyhedral Shapes in Chemistry: 6- and 7-Coordination. ACS Omega. 2025 Oct 14;10(40):47189–209. doi:10.1021/acsomega.5c05878
51. Zadrozny JM, Niklas J, Poluektov OG, Freedman DE. Millisecond Coherence Time in a Tunable Molecular Electronic Spin Qubit. ACS Cent Sci. 2015 Dec 23;1(9):488–92. doi:10.1021/acscentsci.5b00338